\documentclass[twocolumn,aps,superscriptaddress,prb,longbibliography,preprintnumbers]{revtex4-2}

\usepackage{graphicx}
\usepackage{overpic}
\usepackage{float}
\usepackage{amsmath}
\usepackage{amssymb}
\usepackage{bm}
\usepackage{times}
\usepackage{physics}
\usepackage{siunitx}
\usepackage{comment}
\usepackage{hyperref}

\begin{document}
\makeatother

\title{
Non-Abelian Chern band in rhombohedral graphene multilayers
}

\author{Taketo Uchida}
\affiliation{Department of Physics, The University of Osaka, Toyonaka, Osaka 560-0043, Japan}
\author{Takuto Kawakami}
\affiliation{Department of Physics, The University of Osaka, Toyonaka, Osaka 560-0043, Japan}
\author{Mikito Koshino}
\affiliation{Department of Physics, The University of Osaka, Toyonaka, Osaka 560-0043, Japan}

\date{\today}

\begin{abstract}
Moiré flat bands in rhombohedral multilayer graphene provide a platform for exploring interaction-driven topological phases, where a single isolated band often forms a Chern band. 
However, non-Abelian degenerate Chern bands with internal symmetries such as $\mathrm{SU}(N)$ have so far been realized only in highly engineered systems.
Here, we show that a doubly degenerate non-Abelian Chern band with Chern number $|C|=1$ emerges spontaneously at filling $\nu=2$ in rhombohedral 3-, 4-, and 5-layer graphene, regardless of the presence of an hBN substrate. 
Using self-consistent Hartree-Fock calculations, we map out phase diagrams as functions of displacement field and electronic periodicity, and analytically demonstrate that the Fock term drives spontaneous symmetry breaking and generates non-Abelian Berry curvature.
We further show that this non-Abelian topology is characterized by $\mathrm{SU}(2)$ gauge flux threading the noncontractible cycles of the Brillouin zone, leading to a global non-Abelian holonomy.
Our findings unveil a new class of interaction-driven non-Abelian topological phases, distinct from quantum anomalous Hall and fractional Chern phases.
\end{abstract}

\maketitle

{\it Introduction ---} Recent years have witnessed rapid progress in exploring topological states of matter in moir\'e systems, including the observation of integer and fractional quantum anomalous Hall (IQAH and FQAH) states ~\cite{xie2021fractional,cao2018correlated,stepanov2021competing,zondiner2020cascade,park2021flavour,wong2020cascade,choi2021interaction,yu2022correlated,chen2021electrically,polshyn2020electrical,polshyn2022topological,zhang2023local,cao2020tunable,burg2019correlated,liu2020tunable,he2021symmetry,shen2020correlated,xia2025topological,li2021quantum,zhao2024realization,tao2024valley,foutty2024mapping,cai2023signatures,zeng2023thermodynamic,park2023observation,xu2023observation,wu2019topological,yu2020giant,zhai2020theory,tang2021geometric,devakul2021magic,zhang2021spin,wang2023staggered,pan2020band,abouelkomsan2024multiferroicity,crepel2024chiral,li2024electrically,qiu2023interaction,li2021spontaneous,crepel2023anomalous,morales2023pressure,wang2024fractional,reddy2023fractional,jia2024moire,yu2024fractional,wang2023diverse,goldman2023zero,reddy2023toward,dong2023composite,morales2024magic,lu2024fractional,choi2025superconductivity,chen2020tunable,zhou2021half,han2024large,chen2022tunable,spanton2018observation,chen2019signatures,zhang2019nearly,chittari2019gate,repellin2020ferromagnetism,zhang2019bridging}. 
Rhombohedral multilayer graphene has emerged as a particularly promising platform for such studies, owing to its flat bands and enhanced electronic correlations under displacement fields ~\cite{dong2024anomalous,dong2024theory,zhou2024fractional,guo2024fractional,kwan2023moir,zhou2024new,kudo2024quantum,lu2024fractional,choi2025superconductivity,seiler2022quantum,chen2020tunable,zhou2021half,han2024correlated,han2024large,chen2022tunable,spanton2018observation,chen2019signatures,liu2023interaction,han2023orbital,sha2024observation,zhang2019nearly,chittari2019gate,repellin2020ferromagnetism,zhang2019bridging}. 
In particular, quantum Hall crystal states at filling factor $\nu = 1$ have been identified as a fertile ground for novel correlated and topological phenomena, which can occur even in the absence of a moir\'e potential from the hBN substrate ~\cite{dong2024anomalous,dong2024theory,zhou2024fractional,guo2024fractional,kwan2023moir,zhou2024new}.
In these systems, Chern bands typically manifest as single isolated bands with nonzero Berry curvature, whose momentum-space integral yields an integer Chern number. 
More generally, however, the notion of Berry curvature extends to degenerate bands, where it acquires a non-Abelian character described by gauge groups such as SU($N$). 
While non-Abelian Chern bands are of significant theoretical interest, their realization in realistic solid-state materials has remained elusive, with experimental demonstrations thus far limited to highly engineered platforms such as ultracold atomic gases
~\cite{yang2019synthesis,goldman2014light,goldman2007quantum,goldman2009ultracold,goldman2009non,osterloh2005cold,parto2023non,halimeh2024spin,surace2024scalable,essin2012antiferromagnetic,gorshkov2010two,cazalilla2014ultracold,cooper2013reaching}.

In this work, we report a novel quantum phase at $\nu = 2$ 
realized in rhombohedral multilayer graphene.  
This phase is distinct from previously known 
topological phases and is characterized by   
a doubly spin-degenerate band carrying 
a total Chern number of $|C| = 1$ and non-commutative 
$\mathrm{SU}(2)$ gauge structure, 
and a skyrmionic spin texture with a magnetic winding number of 2.  
This unexpected behavior originates from 
the geometric structure of the degenerate band subspace,
 which supports a matrix-valued 
Berry connection due to spin degeneracy.  
As a result, the band topology is 
characterized by a nontrivial $\mathrm{SU}(2)$ holonomy, including $\mathrm{SU}(2)$ gauge flux threading the noncontractible cycles of the Brillouin zone. 
We therefore refer to this phase as a non-Abelian state.

Using self-consistent Hartree-Fock (HF) calculations, we construct phase diagrams for rhombohedral 3-, 4-, and 5-layer graphene by systematically varying the displacement field and the characteristic period of the system, which corresponds either to a moir\'e superlattice period (in the presence of hBN) or to the intrinsic electronic order period (in its absence).
These phase diagrams demonstrate that the non-Abelian phase appears across a wide parameter range.
They also show that this phase competes with other known phases, including the quantum spin Hall (QSH) phase ~\cite{kudo2024quantum} and metallic states, which emerge under different conditions.
Our results indicate that this non-Abelian phase can emerge under experimentally realistic conditions,
making it a compelling target for future experimental observation.

To further elucidate the nature of the non-Abelian Chern bands, we introduce a minimal theoretical model based on a generic parabolic band structure. 
This $2 \times 2$ Hamiltonian captures the essential aspects of the non-Abelian properties and the associated spin texture.  
We show that the model is invariant under a combined symmetry of half-lattice translation and spin rotation at every momentum point in the Brillouin zone (BZ).
As a consequence, the degenerate band doublet acquires noncommuting $\mathrm{SU}(2)$ holonomies upon adiabatic transport along the noncontractible cycles of the BZ, which can be viewed as $\mathrm{SU}(2)$ gauge flux threading the BZ torus.

{\it Single-particle Hamiltonian ---} 
Here we present the single-particle Hamiltonian 
for a rhombohedral $N_L$-layer graphene system aligned with an hBN substrate at a twist angle $\theta$.
We define the moir\'{e} reciprocal lattice vectors based on the lattice mismatch and twist angle between graphene and hBN.  
To isolate the effect of the hBN potential, we also consider a case where the moir\'{e} potential is set to zero while keeping the same reciprocal lattice vectors.
Let the primitive lattice vectors of graphene and the hBN substrate be $\bm{a}_{j}$ and $\bm{a}'_{j}$, respectively. 
The lattice constants of graphene and hBN are 
$a=|\bm{a}_{j}|=0.246$ nm and $a_{\rm hBN} = |\bm{a}'_{j}|=0.2504$ nm, respectively.
These vectors are related as
$\bm{a}'_{i} = MR_{\theta}\bm{a}_{i}$,
where $R_{\theta}$ represents the rotation matrix corresponding to the twist angle $\theta$, and $M = (1+\epsilon)I$ accounts for the lattice mismatch with $\epsilon = a_{\rm hBN}/a - 1 \simeq 1.8\ \%$.
We define the reciprocal lattice vectors 
$\bm{G}_{j}$ and $\bm{G}'_{j}$ for graphene and hBN, satisfying the relations $\bm{a}_{i}\cdot\bm{b}_{j} =\bm{a}'_{i}\cdot\bm{b}'_{j} = 2\pi\delta_{ij}$.
The moir\'e reciprocal lattice vectors are defined as
$ \bm{G}^M_{j} = \bm{b}_{j} - \bm{b}'_{j}$.

The continuum Hamiltonian of $N_L$-layer rhombohedral graphene with hBN is given by \cite{koshino2009trigonal,zhou2021half,dong2024anomalous,dong2024theory,zhang2010band,jung2013gapped}
\begin{align}
    \hat{h}^{(N_L)}_{\rm{RG}}(\bm{k}) =
    \begin{pmatrix}
        h_1 + V_{\rm{hBN}} & f & g & &\\
        f^{\dag} & h_2 & f & g & \\
        \ddots & \ddots & \ddots & \ddots & \ddots\\
        & g^{\dag} & f^{\dag} & h_{(N_L-1)} & f \\
        & & g^{\dag} & f^{\dag} & h_{N_L}
    \end{pmatrix},
    \label{single-particleH}
\end{align}
where
\begin{align}
    h_l &=
    \begin{pmatrix}
        0 & v_0 k_- \\
        v_0 k_+ & 0
    \end{pmatrix}
    + 
    \begin{pmatrix}
    u_D (l - 1) & 0 \\
        0 & u_D (l - 1)
    \end{pmatrix},
    \\
    f &=
    \begin{pmatrix}
        v_{4} k_+ & v_{3} k_-  \\
        t_1& v_{4} k_+
    \end{pmatrix},\ 
    g =
    \begin{pmatrix}
        0 & t_2/2 \\
        0 & 0
    \end{pmatrix}.  
\end{align}
The basis of this Hamiltonian corresponds to the $N_L$-layer graphene sublattices 
$(A_1,B_1,A_2,B_2,\cdots,A_{N_L},B_{N_L})$.
The potential from hBN substrate, which is in  contact with the first graphene layer, affects only the $A_1$ and $B_1$ sublattices. The moir\'e potential $V_{\rm{hBN}}$ can be written as \cite{moon2014electronic}:
\begin{align}
    V_{\rm{hBN}} &= V_0
    \begin{pmatrix}
        1 & 0\\
        0 & 1
    \end{pmatrix} \notag
    \\ \notag
    &+ \bigg\{
    V_1e^{i\xi\psi}
    \bigg[
    \begin{pmatrix}
        1 & \omega^{-\xi}\\
        1 & \omega^{-\xi}
    \end{pmatrix}
    e^{i\xi\bm{G}^M_1\cdot\bm{r}}
    +\begin{pmatrix}
        1 & \omega^{\xi}\\
        \omega^{\xi} & \omega^{-\xi}
    \end{pmatrix}
    e^{i\xi\bm{G}^M_2\cdot\bm{r}}\\ 
    &+
    \begin{pmatrix}
        1 & 1\\
        \omega^{-\xi} & \omega^{-\xi}
    \end{pmatrix}
    e^{-i\xi(\bm{G}^M_1+\bm{G}^M_2)\cdot\bm{r}}
    \bigg]
    +\rm{H.c.}\bigg\},
\end{align}
where $\omega=e^{2\pi i/3}$ and the parameters $(V_0,V_1,\psi) = (28.9\ \rm{meV},21.0\ \rm{meV}, -0.29\ \rm{rad})$ \cite{moon2014electronic}.
Under the continuum approximation, the parameters $v_i$ and the momenta $k_{\pm}$ are given by $v_i = \sqrt{3}a t_i/2$ and $k_{\pm} = \xi k_x \pm ik_y$.
where $\xi = +1\ (-1)$ denotes the valley $K\ (K')$.
Here, $t_0$ and $(t_1,t_2,t_3,t_4)$ represent the intralayer and interlayer hopping parameters, while $u_D$ denotes the interlayer potential difference induced by the perpendicular displacement field.
In our calculations, we adopt the parameter set 
$(t_0,t_1,t_2,t_3,t_4)
=(3100,380,-21,290,141)\ \rm{meV}$~\cite{zibrov2018emergent}.
When $u_D > 0$, the topological surface states of rhombohedral multilayer graphene exhibit valence bands composed of electrons on the moir\'{e}-proximate side and conduction bands composed of electrons on the moir\'{e}-distant side. 
In this paper, we focus on the conduction bands for the moir\'{e}-distant side.

{\it Hartree-Fock calculation---} 
We employ the self-consistent Hartree–Fock (HF) method to describe the effects of electron–electron interactions in rhombohedral multilayer graphene \cite{dong2024anomalous,bultinck2020ground}, 
with the derivation detailed in Ref.~\cite{SM}.
Let $E_{\bm{k}\alpha}$ and $\psi_{\bm{k}\alpha}$ denote the eigenenergy and eigenstate of the single-particle Hamiltonian $h_0(\bm{k})$, where $\alpha$ is an index representing the band, spin, and valley degrees of freedom.  
We define $c^{\dag}_{\bm{k}\alpha}$ as the creation operator for the single-particle eigenstate $\psi_{\bm{k}\alpha}$.  
The Hartree–Fock Hamiltonian is written as  
$ h(\bm{k}) = h_{\rm{0}}(\bm{k})+ h_{\rm{H}}(\bm{k})+ h_{\rm{F}}(\bm{k})$,
where
\begin{align}
    h_{\rm{H}}(\bm{k})
    &=\frac{1}{A}
    \sum_{\bm{G}^M}
    V_{\bm{G}^M}
    \Lambda_{\bm{G}^M}(\bm{k})
    \sum_{\bm{k}'}
    \operatorname{Tr}[
    P(\bm{k}')
    \Lambda_{\bm{G}^M}(\bm{k}')^*
    ],
\nonumber \\
    h_{\rm{F}}(\bm{k})
    &=-\frac{1}{A}
    \sum_{\bm{q}}
    V_{\bm{q}}
    \Lambda_{\bm{q}}(\bm{k})
    P^T(\bm{k}+\bm{q})
    \Lambda_{\bm{q}}(\bm{k})^{\dag}.
    \label{eq_Fock_form}
\end{align}
Here, $A$ denotes the system area, and $\bm{G}^M = n_1 \bm{G}^M_1 + n_2 \bm{G}^M_2$ runs over moir\'e reciprocal lattice vectors, where $n_1, n_2$ are integers. The $V_{\bm{q}}$ is the Fourier transform of the gate-screened Coulomb interaction, given by
\begin{align}
    V_{\bm{q}}&=\frac{e^2}{2\epsilon_0\epsilon_r |\bm{q}|}\tanh{(|\bm{q}|d)},
\end{align}
where we assume a relative dielectric constant $\epsilon_r = 5$ and a gate separation $d = 25\,\mathrm{nm}$ in this study.  
The single-particle density matrix $P(\bm{k})$ and the form factor $\Lambda_{\bm{q}}(\bm{k})$ are written in the basis of single-particle eigenstates as  
\begin{align}
[P(\bm{k})]_{\alpha\beta}
&=\ev*{c_{\bm{k}\alpha}^{\dag}c_{\bm{k}\beta}},
\\
 [\Lambda_{\bm{q}}(\bm{k})]_{\alpha\beta}
&=\bra{\psi_{\bm{k}\alpha}}
e^{-i\bm{q}\cdot\bm{r}}
\ket{\psi_{\bm{k}+\bm{q}\beta}},
\end{align}
where $\langle \cdots \rangle$ denotes the expectation value with respect to the many-body ground state at a given electron filling.
The total energy $E_{\rm{tot}}$ is evaluated as
\begin{align}
    E_{\rm{tot}}=
    \frac{1}{A}
    \sum_{\bm{k}}
    \operatorname{Tr}\bigg[
    \bigg(h_{0} (\bm{k})+ 
    \frac{h_{\rm{H}}(\bm{k})+h_{\rm{F}}(\bm{k})}{2}\bigg)^{\rm{T}} P(\bm{k})
    \bigg].
\end{align}

The Hartree term $h_{\rm{H}}(\bm{k})$, the Fock term $h_{\rm{F}}(\bm{k})$, the density matrix $P(\bm{k})$, and the form factors $\Lambda_{\bm{q}}(\bm{k})$ are all defined within the first moiré BZ and periodic with respect to $\bm{k}$. 
In contrast, the wave vector $\bm{q}$ of $\Lambda_{\bm{q}}(\bm{k})$ is not restricted to the first BZ; it must be specified for each $\bm{q}$ vector across distant BZs.
In our numerical implementation, we divide each moir\'{e} BZ into a uniform $24 \times 24$ $\bm{k}$-point mesh. 
In Eq.~\eqref{eq_Fock_form}, the summation over $\bm{G}^M$ is restricted to 19 moiré BZs defined by $\bm{G}^M = n_1 \bm{G}^M_1 + n_2 \bm{G}^M_2$ with $|n_1|, |n_2| \leq 2$, while the summation over $\bm{q}$ is carried out for each $\bm{q}$ vector at every $\bm{k}$-point in the $24 \times 24$ mesh covering all these zones.
For the Hartree-Fock calculation, we include the lowest 7 conduction bands for each spin and valley (Ref.~\cite{SM}).
To explore the ground state at filling $\nu=2$, we initialize $P(\bm{k})$ with small random complex numbers.
Two electrons are assigned per assumed unit cell, either both in the same valley or one in each of the $K$ and $K'$ valleys.
The self-consistent iterations are continued until the electron density matrix $P(\bm{k})$ converges.

\begin{figure}[t]
    \centering
    \includegraphics[width=\linewidth]{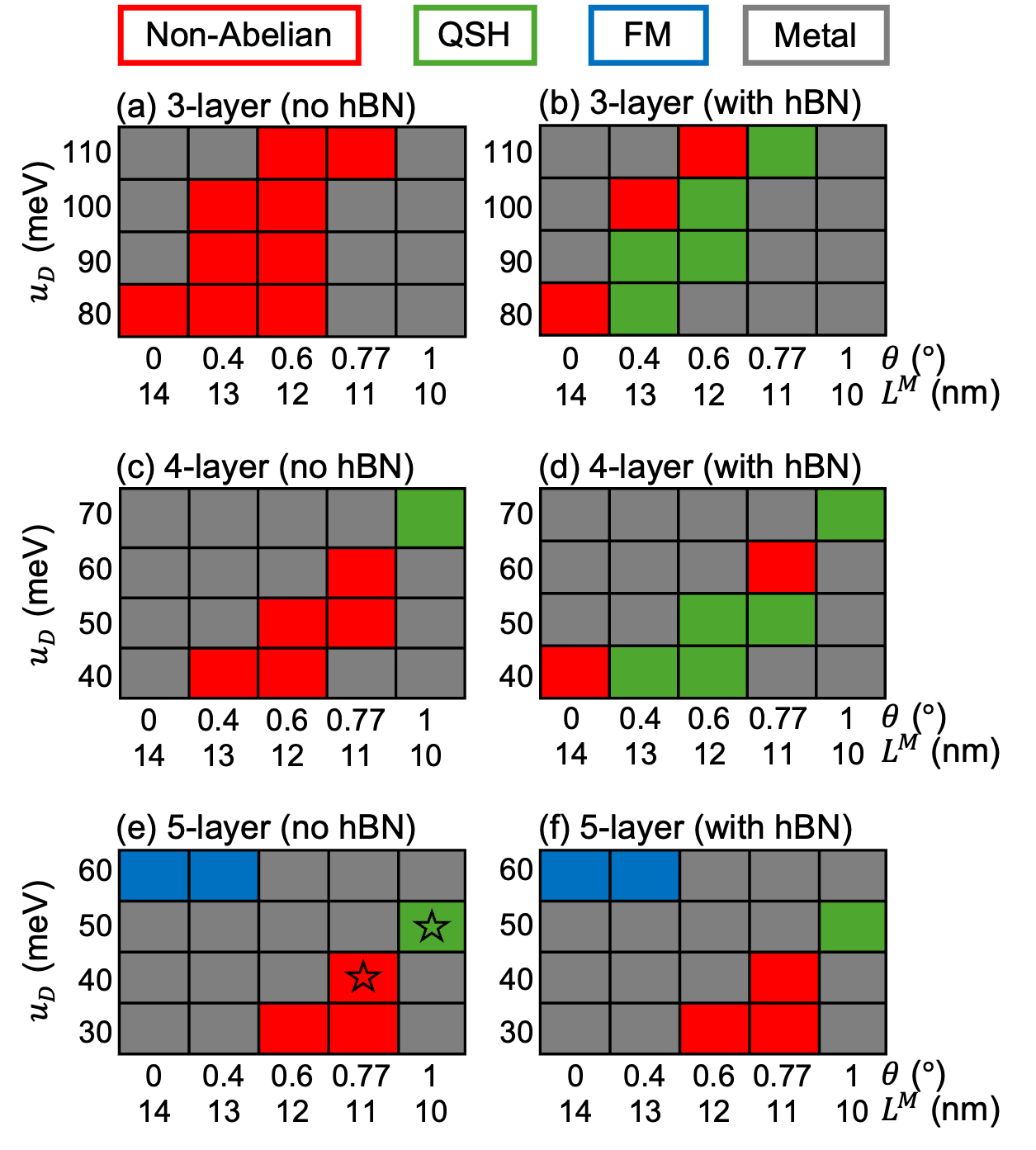}
\begin{flushleft}
    \caption{
Phase diagram of rhombohedral multilayer graphene at filling factor $\nu = 2$  
as a function of the interlayer potential difference $u_D$ and the twist angle $\theta$ (with corresponding moiré period $L^M$).  
The upper, middle, and bottom rows correspond to $N_L = 3$, $4$, and $5$, respectively, and the left and right panels show results without and with $V_{\rm{hBN}}$, respectively.  
Gray, green, and red regions indicate the metallic phase, the QSH phase, and the non-Abelian phase, respectively.
  }
\label{fig:PhaseDiagram}
\end{flushleft}
\end{figure}

\begin{figure}[t]
    \begin{center}
    \includegraphics[width=1\linewidth]{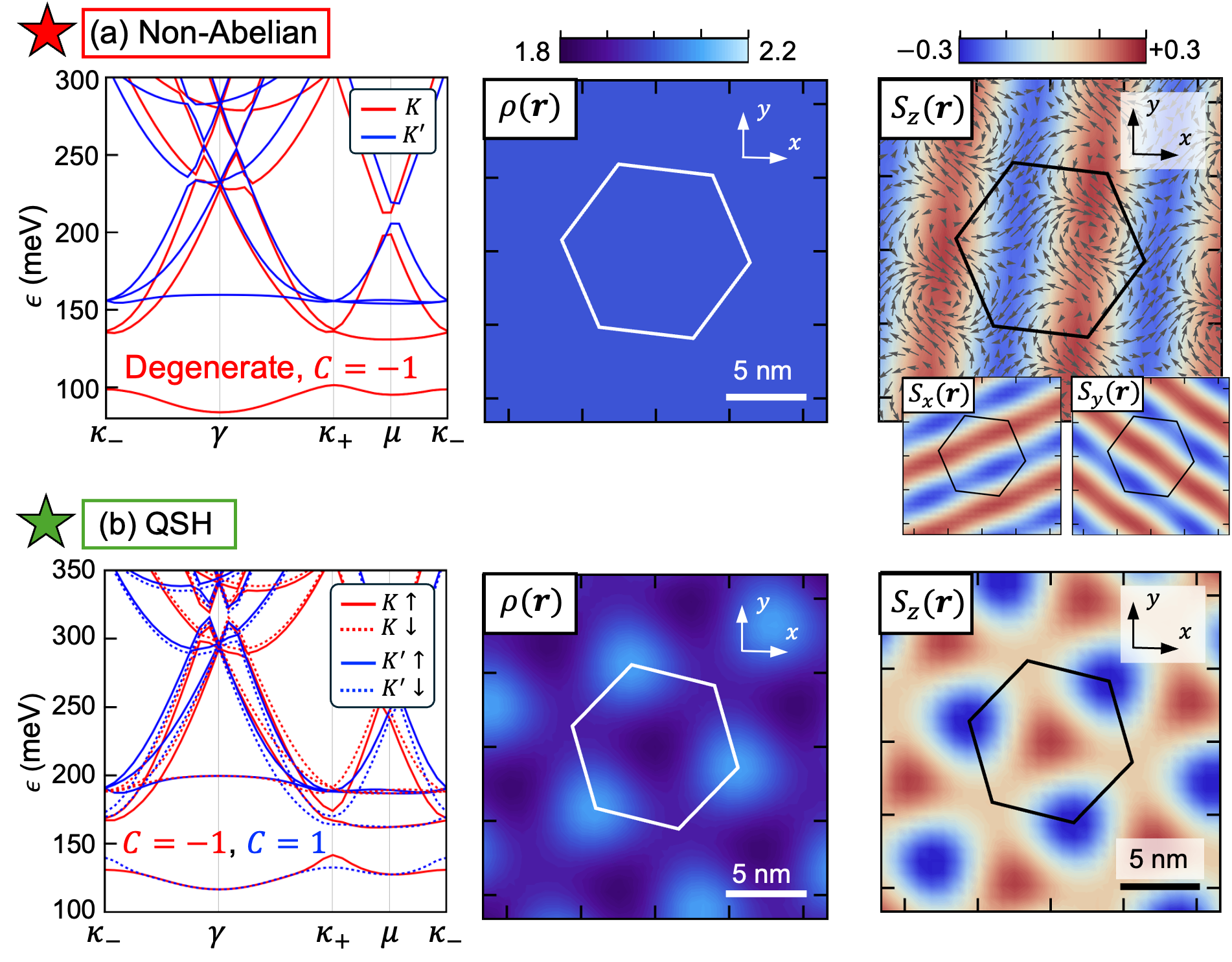}
    \end{center}
\begin{flushleft}
  \caption{
  Band structure, local charge density and local spin texture  of (a) the non-Abelian state and (b) the QSH state obtained by HF calculations for the 5-layer graphene, which are marked by red and green stars, respectively, in Fig.~\ref{fig:PhaseDiagram}.
  The hexagon in the middle and right figures represent the superlattice unit cell.
    In the right panel, the color map represents 
    the local spin-density $S_z$ while the arrow represents $S_x$ and $S_y$ components.
    }
  \label{fig:NA_QSH}
\end{flushleft}
\end{figure}

{\it Ground states at} $\nu=2$ --- Using self-consistent HF calculations, we compute the ground states of rhombohedral $N_L$-layer graphenes at filling factor $\nu = 2$, for various displacement fields $u_D$ and twist angles $\theta$.  
Figure \ref{fig:PhaseDiagram} summarizes the results: the upper, middle, and bottom rows correspond to $N_L = 3$, $4$, and $5$, respectively.  
In each row, the left and right panels show the phase diagrams without and with $V_{\rm{hBN}}$, respectively.
The calculation without $V_{\rm{hBN}}$ corresponds to a situation in which the moiré unit cell is fixed while the moiré potential from hBN is simply turned off.
In these phase diagrams, the gray regions represent metallic phases, where charge and spin orders are absent and the energy spectrum remains ungapped.  
The green regions indicate the quantum spin Hall (QSH) phase, which was also reported in a previous study~\cite{kudo2024quantum}.  
The red regions correspond to a non-Abelian phase, newly identified in this work and described in detail below.
The blue regions represent ferromagnetic (FM) phases characterized by a valley- and spin-polarized anomalous Hall crystal. 
In FM phases, the Chern bands are non-degenerate, and the total Chern number is $|C|=1$ ~\cite{dong2024anomalous,zhou2024new}.

In Fig.~\ref{fig:NA_QSH}, we compare the band structure (left), local charge density (middle), and local spin texture (right)  
for (a) the non-Abelian state and (b) the QSH state in the 5-layer system without $V_{\rm{hBN}}$,  
which are marked by red and green stars, respectively, in Fig.~\ref{fig:PhaseDiagram}.
The non-Abelian phase [Fig.~\ref{fig:NA_QSH}(a)] is an integer Chern insulator characterized by two completely spin-degenerate bands with a single valley sector.
The Chern number for these degenerate bands can be evaluated by integrating the $2\times 2$ matrix-valued non-Abelian Berry curvature over the BZ \cite{fukui2005chern}, and that for the filled band doublet is found to be $C=-1$.
If the two bands are filled in the opposite valley $K'$ instead, the total Chern number and the magnetic winding number both flip their signs.
The charge density $\rho(\bm{r})$ in the middle panel shows a nearly uniform distribution. 
In the rightmost panel, the color map represents 
the spin-density component
$S_z$, while the arrow represents $S_x$ and $S_y$.
The $S_x$,  $S_y$ and  $S_z$ exhibit sinusoial modulations
along three directions with 120$^\circ$ rotations.
As a whole,the spin texture exhibits a skyrmionic distribution with a magnetic winding number of $+2$.
Here, we globally rotate the spin axes in an appropriate manner to reveal the symmetric stripe patterns in $S_x$, $S_y$, and $S_z$ shown in Fig.~\ref{fig:NA_QSH}(a).

In contrast, the QSH state [Fig.~\ref{fig:NA_QSH}(b)] has one spin-polarized Chern band in each valley with $C = \pm 1$, resulting in a total Chern number of zero. It exhibits significant spatial modulation in the charge density and a collinear spin texture with $S_x=S_y=0$.  
This corresponds to a superposition of two anomalous Hall crystals with $|C| = 1$: one with spin up and the other with spin down~\cite{dong2024anomalous}.  
In the present calculation, the QSH state is degenerate in energy with the QAH state, which consists of two valley-polarized $|C| = 1$ bands.  
It was shown that the QSH state is slightly more stable than the QAH state~\cite{kudo2024quantum}.

Without $V_{\rm{hBN}}$, the non-Abelian phase appears over a wide range of parameter space in the 3-, 4-, and 5-layer systems (see Fig.~\ref{fig:PhaseDiagram}).  
With $V_{\rm{hBN}}$, the QSH state becomes more dominant in the 3- and 4-layer cases, while in the 5-layer case, the phase diagram remains largely unaffected.  
This is presumably because the occupied conduction bands are sufficiently distant from the hBN substrate.  
In the presence of $V_{\rm{hBN}}$, the degeneracy of the Chern bands in the non-Abelian state is slightly lifted by a few meV,  
although the charge and spin density profiles remain nearly unchanged 
(Ref.~\cite{SM}).

\begin{figure}[t]
    \begin{center}
    \includegraphics[width=1.\linewidth]{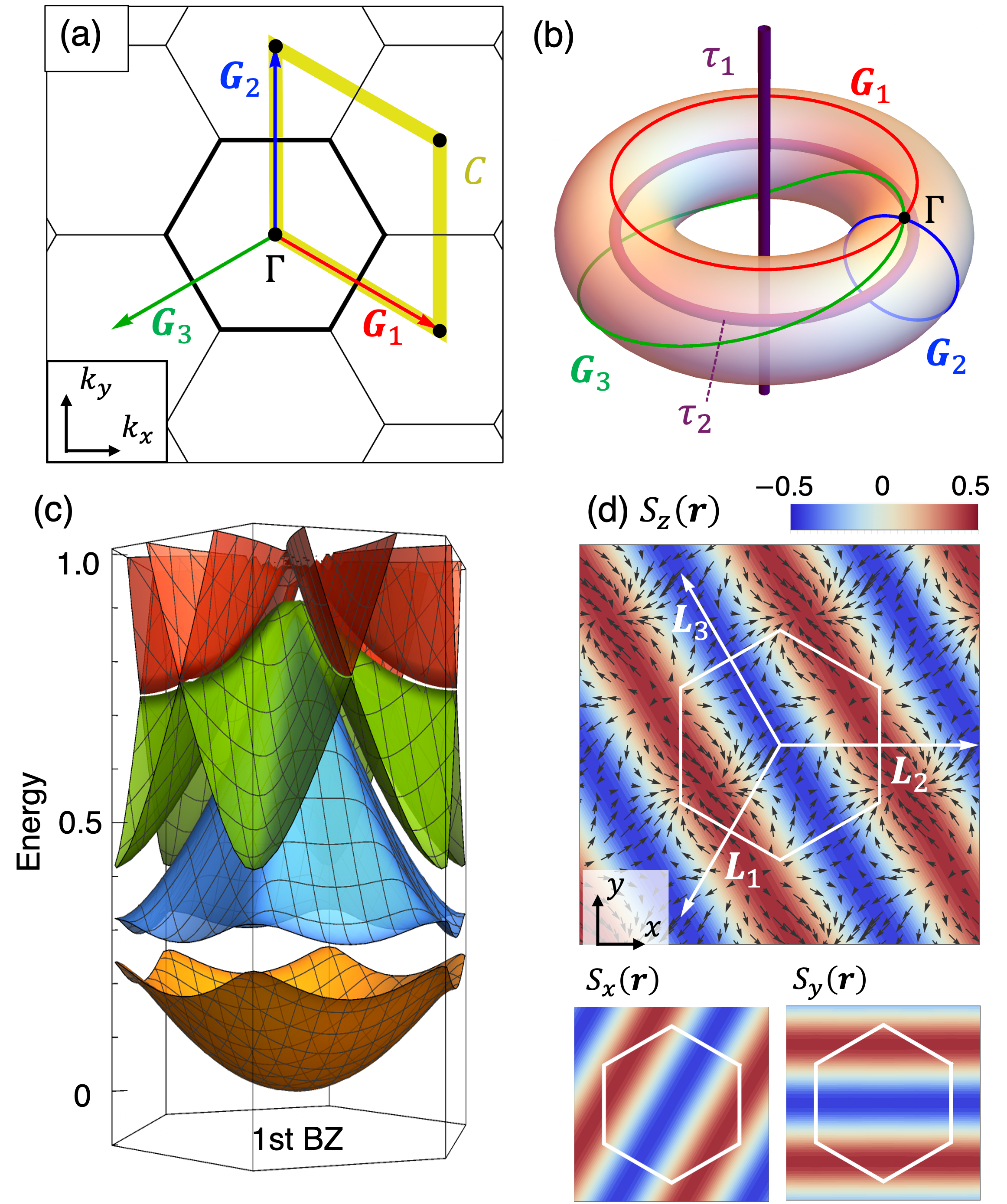}
    \end{center}
\begin{flushleft}
 \caption{
(a) BZ of the model Hamiltonian in Eq.~\eqref{eq_H_simple} with reciprocal vectors $\bm{G}_i$.
(b) BZ torus with noncontractible cycles corresponding to paths from $\Gamma$ to $\bm{G}_i$ $(i=1,2,3)$.
Thick lines indicate $\mathrm{SU}(2)$ gauge fluxes associated with $\tau_1$ (vertical line) and $\tau_2$ (circular loop inside the torus).
(c) Band structure for the parameter $L=1$, $\hbar^2/(2m)=1$ and $V_0=0.05$. Each band is two-fold degenerate on the whole BZ.
(d) Local spin density When the lowest band doublet is fully occupied, plotted in the same manner as in Fig.~\ref{fig:NA_QSH}.
}
  \label{fig:eff_model}
\end{flushleft}
\end{figure}

{\it Simple model arguments ---} 
The characteristics of the non-Abelian phase---its twofold spin degeneracy with Chern number $|C| = 1$ and skyrmion-like spin texture---can be understood using a simple two-dimensional model Hamiltonian with a spin-dependent potential that approximates the mean-field HF potential.
The model is explicitly expressed as,
\begin{align}
    &H = \frac{\bm{p}^2}{2m}
  + 2V_0 \sum_{i=1}^3
  \sigma_i \cos \bm{G}_i\cdot\bm{r}
    \label{eq_H_simple}
\end{align}
where 
$\bm{p} = (p_x,p_y)$ is the momentum,
$\sigma_i$ is the Pauli matrices, and
$\bm{G}_i = G(\cos \theta_i, \sin \theta_i)$ with 
$\theta_i = 2\pi(i-1)/3 - \pi/6 (i=1,2,3)$ 
are a set of trigonally symmetric wave vectors [see, Fig.~\ref{fig:eff_model}(a)].
Accordingly, the system is translationally symmetric with the lattice vectors 
$\bm{L}_i = L(\cos (\theta_i-\pi/2), \sin (\theta_i-\pi/2))\,(i=1,2,3)$ where $L=4\pi/(\sqrt{3}G)$.
Note that $\bm{L}_i\cdot\bm{G}_j$ is equal to $0, \pm 2\pi$ 
when $i-j \equiv 0, \pm 1$ in modulo 3, respectively.

In Fig.~\ref{fig:eff_model}(c), we show the band structure for the parameters $L = 1$, $\hbar^2/(2m) = 1$, and $V_0 = 0.05$.  
The parabolic band is folded into a hexagonal BZ, with a band gap opening at the zone boundary.  
The resulting energy bands share the same essential properties as the non-Abelian phase in rhombohedral graphene multilayers [Fig.~\ref{fig:NA_QSH}(a)]: 
each band is twofold degenerate throughout the BZ, and the Chern number of the lowest band doublet is $C = 1$.  
When the lowest band doublet is fully occupied, the resulting spin density distribution, shown in Fig.~\ref{fig:eff_model}(b),  
exhibits a skyrmionic texture with a winding number of $+2$, similar to Fig.~\ref{fig:NA_QSH}(a).  
The density plots of $S_x$, $S_y$, and $S_z$ display stripe patterns along three directions related by 120$^\circ$ rotations, consistent with the symmetric form of the Hamiltonian.

The twofold band degeneracy can be explained by a non-symmorphic symmetry, expressed by $[H, U_i]=0$ where
\begin{align}
    & U_i = \sigma_i T_{\bm{L}_i/2} \, (i=1,2,3),
    \label{eq_U}
\end{align}
and $T_{\bm{R}}$ represents the translation by a vector $\bm{R}$.
Because of the property of the Pauli matrices, the operators $U_i$'s anticommute each other, and it leads to two-fold degeneracy for all the bands at any Bloch wave number $\bm{k}$.
The eigenvalues of $U_i$ for the two degenerate states at $\bm{k}$ is given by $\pm e^{i\bm{k}\cdot\bm{L}_i/2}$,
because the eigenvalue of $U_i^2 = T_{\bm{L}_i}$ is $e^{i\bm{k}\cdot\bm{L}_i}$.
Therefore, when we continuously move the wave number $\bm{k}$ by a primitive reciprocal lattice vector $\bm{G}_j$, the eigenvalues of $U_i$ for the degenerate states is swapped when $i \neq j$, because $e^{i\bm{G}_j\cdot\bm{L}_i/2} = -1$.
This indicates that the degenerate energy bands cannot be separated into two independent sectors by the eigenvalues of any $U_i$.

The eigenvalue flipping of $U_i$ can be interpreted as the presence of a non-Abelian $\mathrm{SU}(2)$ gauge flux.  
For the lowest-energy band doublet at each $\bm{k}$, we can fix the $\mathrm{SU}(2)$ gauge such that $U_i$ $(i=1,2,3)$ is identified with $\tau_i$, where $\tau_i$ are the Pauli matrices acting in the doublet space.  
When a Bloch state within the doublet is adiabatically transported across the BZ from the $\Gamma$ point to $\bm{G}_i$, one finds that the initial wavefunction $\psi$ is transformed into
$\psi' = - \tau_i \psi = i e^{i\pi\tau_i/2}\psi$ at the final point.  
That is, the $\tau$ pseudospin undergoes a $\pi$ rotation about the $\tau_i$ axis, corresponding to a swapping of the eigenvalues of $U_i$.
This behavior indicates that two distinct types of $\mathrm{SU}(2)$ gauge flux thread the noncontractible cycles of the BZ torus, as schematically illustrated in Fig.~\ref{fig:eff_model}(b).  
Here, the cycles in the toroidal and poroidal directions correspond to translations by $\bm{G}_1$ and $\bm{G}_2$, respectively,  
which encircle the gauge fluxes associated with $\tau_1$ and $\tau_2$ rotations.  
A translation by $\bm{G}_3$ corresponds to a loop encircling both fluxes and results in a $\tau_3 (\propto \tau_1\tau_2)$ rotation.

The Chern number of a degenerate band can be obtained by evaluating the Berry phase along a closed path $C$ 
around the BZ [Fig.~\ref{fig:eff_model}(b)].
Upon completing this closed path, the initial state $\psi$ is transformed as
$\psi' = (-\tau_2) (-\tau_1)(-\tau_2) (-\tau_1) \psi = - \psi$, leaving a $\mathrm{U}(1)$ phase of $\pi$ with no residual $\mathrm{SU}(2)$ rotation.
Therefore, the total phase factor accumulated by the degenerate band doublet is $\pi\times 2 = 2\pi$, yielding a Chern number $C = 1$.
Note that no $\mathrm{SU}(2)$ flux penetrates the surface of the Brillouin-zone torus; consequently, only a $\mathrm{U}(1)$ phase is accumulated when enclosing contractible cycles.

A global $\mathrm{SU}(2)$ holonomy associated with nontrivial loops in the Brillouin zone signals genuinely non-Abelian band topology.
A natural transport probe is to weakly dope the $\nu=2$ insulator so that a small carrier density occupies the non-Abelian Chern doublet.
Assuming the ordered state persists, an external electric field drives semiclassical motion in momentum space causing the internal spinor to acquire a path-ordered $\mathrm{SU}(2)$ rotation. This enables a non-Abelian Aharonov–Bohm effect~\cite{yang2019synthesis} detectable through interference between wave packets along distinct trajectories, and 
Bloch oscillations~\cite{di2020non} along nontrivial trajectories lead to path-dependent $\mathrm{SU}(2)$ rotations.
Moreover, the matrix-valued Berry connection may influence optical transition matrix elements across the gap, together with the associated polarization selection rules, providing an additional probe of the non-Abelian band geometry~\cite{yang2014nonlinear}.
Finally, consistent with bulk–edge correspondence, the Chern insulator hosts in-gap edge states. A systematic analysis of the edge-state structure and its connection to the bulk non-Abelian band geometry remains an important direction for future research.

We note that the present analysis is based on self-consistent Hartree–Fock theory, which is expected to capture the formation of interaction-driven ground states but may overestimate the stability of ordered phases~\cite{kwan2023moir,xie2020nature,bultinck2020ground,hejazi2021hybrid,lian2021twisted,parker2021field,liu2021nematic,kwan2022skyrmions,faulstich2023interacting,dong2024anomalous,dong2024theory,zhou2024fractional}.
In rhombohedral multilayer graphene at $\nu=1$, previous studies have shown that correlation effects beyond Hartree–Fock can suppress crystalline anomalous Hall states~\cite{yu2025moire}, while alignment with an hBN substrate may play an important role in stabilizing such phases~\cite{kwan2025moire,huo2025does}.
A systematic investigation of the stability of the non-Abelian phase beyond Hartree–Fock theory will be addressed in future studies.

{\it Conclusion ---}
In conclusion, using self-consistent Hartree-Fock calculations in the absence of magnetic fields, we have identified a novel quantum phase at filling factor $\nu=2$ in rhombohedral 3-, 4-, and 5-layer graphene.
This phase is a spin-degenerate Chern insulator with total Chern number $|C|=1$, which we refer to as the non-Abelian phase, and it emerges robustly irrespective of the presence of an hBN substrate.
We constructed phase diagrams as functions of the external displacement field and the moir\'e (or intrinsic electronic order) periodicity, demonstrating that the non-Abelian phase competes with conventional quantum spin Hall and metallic states, yet remains stable over a wide and experimentally accessible parameter range.
In addition, this phase exhibits a real-space skyrmionic spin texture with magnetic winding number two.
To elucidate the origin of the non-Abelian phase, we introduced a minimal effective single-particle model that captures the essential physics of the degenerate Chern band.
This model analytically demonstrates that both the Chern number and the associated non-Abelian $\mathrm{SU}(2)$ gauge structure are fixed by symmetry, independent of microscopic band details.

Our results establish rhombohedral multilayer graphene as a promising platform for exploring experimentally observable non-Abelian responses in condensed matter systems.
Further analytical understanding of the microscopic mechanisms stabilizing this phase remains an important direction for future work.

{\it Acknowledgment ---}
This work was supported in part by JSPS KAKENHI Grants No. JP25K00938, No. JP21H05236, No. JP21H05232, No. JP24K06921 and by JST CREST Grant No. JPMJCR20T3, Japan.

\bibliographystyle{apsrev4-2}

\bibliography{reference}

\clearpage

\onecolumngrid
\section{Details of the Hartree-Fock calculation}
In this section, we derive the self-consistent Hartree-Fock (HF) equations ~\cite{dong2024anomalous,bultinck2020ground}. 
Before applying the mean-field approximation, 
the interaction Hamiltonian is given by
\begin{align}
    \hat{H}_{\rm{int}}=
    \frac{1}{2}
    \int\dd\bm{r}_1\dd\bm{r}_2V(|\bm{r}_1-\bm{r}_2|)
    \hat{\psi}^\dag(\bm{r}_1)\hat{\psi}^\dag(\bm{r}_2)
    \hat{\psi}(\bm{r}_2)\hat{\psi}(\bm{r}_1).
\end{align} 
Applying the mean-field approximation, 
we decompose the interaction Hamiltonian into Hartree and Fock terms
as follows:
\begin{align}
    &\hat{H}_{\rm{int}}
    \simeq \hat{H}_{\rm{H}} + \hat{H}_{\rm{F}},
\end{align}
where
\begin{align}
    &\hat{H}_{\rm{H}}
     = \int\dd\bm{r}_1\dd\bm{r}_2
     V(|\bm{r}_1-\bm{r}_2|)
    \hat{\psi}^{\dag}(\bm{r}_1)\hat{\psi}(\bm{r}_1)
    \ev*{
    \hat{\psi}^{\dag}(\bm{r}_2) \hat{\psi}(\bm{r}_2)
    },
    \\
    &\hat{H}_{\rm{F}}
     = -\int\dd\bm{r}_1\dd\bm{r}_2
     V(|\bm{r}_1-\bm{r}_2|)
    \hat{\psi}^{\dag}(\bm{r}_1)\hat{\psi}(\bm{r}_2)
    \ev*{
    \hat{\psi}^{\dag}(\bm{r}_2) \hat{\psi}(\bm{r}_1)
    }.
\end{align}
The electron-electron interaction $V(\bm{r})$
is given by its Fourier transform:
\begin{align}
    V(|\bm{r}_1-\bm{r}_2|)
    &=\frac{1}{A}\sum_{\bm{q}}V_{\bm{q}}e^{i\bm{q}\cdot(\bm{r}_2-\bm{r}_1)},
\end{align}
where
\begin{align}
    V_{\bm{q}}&=\frac{e^2}{2\epsilon_0\epsilon_r |\bm{q}|}\tanh{(|\bm{q}|d)}.
\end{align}
Here, 
$A$ denotes the system area, and 
$\bm{q}$ represents arbitrary wave vectors.
we adopt the dual gate-screened Coulomb interaction, 
assuming the dielectric constant of $\epsilon_r=5$ 
and a gate separation of $d=25$ nm in this study.
The field operator is defined as
\begin{align}
    \hat{\psi}(\bm{r}_1)&
    =\sum_{\bm{k},\alpha}\psi_{\bm{k}\alpha}(\bm{r}_1)c_{\bm{k}\alpha},
\end{align}
where $\alpha$ represents band, spin, and valley indices.
The single-particle eigenstates satisfy
\begin{align}
    \hat{h}^{(N_L)}_{\rm{RG}}(\bm{k})
    \ket{\psi_{\bm{k}\alpha}}
    &= E^{(0)}_{\bm{k}\alpha} \ket{\psi_{\bm{k}\alpha}},
    \\
    c^{\dag}_{\bm{k}\alpha}\ket{0} &= \ket{\psi_{\bm{k}\alpha}},
\end{align}
where $\hat{h}^{(N_L)}_{\rm{RG}}(\bm{k})$ is the single-particle Hamiltonian defined in Eq.
2 in the main text
, and $c^{\dag}_{\bm{k}\alpha}$ is the creation operator corresponds to the eigenstates $\ket{\psi_{\bm{k}\alpha}}$. 
Integrating the eigenstates over spatial coordinates, we obtain
\begin{align}
     \int\dd\bm{r}
    \psi_{\bm{k}_1\alpha}^*(\bm{r})
    e^{-i\bm{q}\cdot\bm{r}}
    \psi_{\bm{k}_2\beta}(\bm{r})=
    \bra{\psi_{\bm{k}_1\alpha}}
    e^{-i\bm{q}\cdot\bm{r}}
    \ket{\psi_{\bm{k}_2\beta}}.
\end{align}
This integral is non-zero only when 
$\bm{k}_2=\bm{k}_1+\bm{q}$.
Using this result, 
the HF Hamiltonian takes the form 
\begin{align}
    \hat{H}_{\rm{H}}
    &=\frac{1}{A}
    \sum_{\bm{k}_{1}\bm{k_2}}
    \sum_{\alpha,\beta,\alpha',\beta'}
    \sum_{\bm{q}}
    V_{\bm{q}}
    \bra{\psi_{\bm{k}_1\alpha}}
    e^{-i\bm{q}\cdot\bm{r}_1}
    \ket{\psi_{\bm{k}_1+\bm{q}\beta}}
    \ev*{
    c_{\bm{k}_2\alpha'}^{\dag}c_{\bm{k}_2-\bm{q}\beta'}
    }
    \bra{\psi_{\bm{k}_2\alpha'}}
    e^{i\bm{q}\cdot\bm{r}_2}
    \ket{\psi_{\bm{k}_2-\bm{q}\beta'}}
    c_{\bm{k}_1\alpha}^{\dag}c_{\bm{k}_1+\bm{q}\beta}.
\end{align}
Similarly, the Fock term is given by
\begin{align}
    \hat{H}_{\rm{F}}
    &=\frac{1}{A}
    \sum_{\bm{k}_{1}\bm{k_2}}
    \sum_{\alpha,\beta,\alpha',\beta'}
    \sum_{\bm{q}}
    V_{\bm{q}}
    \bra{\psi_{\bm{k}_1\alpha}}
    e^{-i\bm{q}\cdot\bm{r}_1}
    \ket{\psi_{\bm{k}_1+\bm{q}\beta}}
    \ev*{
    c_{\bm{k}_2+\bm{q}\alpha'}^{\dag}c_{\bm{k}_1+\bm{q}\beta}
    }
    \bra{\psi_{\bm{k}_2+\bm{q}\alpha'}}
    e^{i\bm{q}\cdot\bm{r}_2}
    \ket{\psi_{\bm{k}_2\beta'}}
    c_{\bm{k}_1\alpha}^{\dag}c_{\bm{k}_2\beta}.
\end{align}
By assuming that the mini-Brillouin zone due to the moir\'e period 
is sufficiently small and that the wave vectors involved 
in expectation values differ only by reciprocal lattice vectors
$\bm{G}^M$, 
the Hamiltonian becomes
\begin{align}
    \hat{H}_{\rm{H}}
    &=\frac{1}{A}
    \sum_{\bm{k}_1,\bm{k}_2}
    \sum_{\alpha,\beta,\alpha',\beta'}
    \sum_{\bm{G}^M}
    V_{\bm{G}^M}
    \bra{\psi_{\bm{k}_1\alpha}}
    e^{-i\bm{G}^M\cdot\bm{r}_1}
    \ket{\psi_{\bm{k}_1\beta}}
    \ev*{
    c_{\bm{k}_2\alpha'}^{\dag}c_{\bm{k}_2\beta'}
    }
    \bra{\psi_{\bm{k}_2\alpha'}}
    e^{i\bm{G}^M\cdot\bm{r}_2}
    \ket{\psi_{\bm{k}_2\beta'}}
    c_{\bm{k}_1\alpha}^{\dag}c_{\bm{k}_1\beta},
    \\
    \hat{H}_{\rm{F}}
    &=-\frac{1}{A}
    \sum_{\bm{k}}
    \sum_{\alpha,\beta,\alpha',\beta'}
    \sum_{\bm{q}}
    V_{\bm{q}}
    \bra{\psi_{\bm{k}\alpha}}
    e^{-i\bm{q}\cdot\bm{r}_1}
    \ket{\psi_{\bm{k}+\bm{q}\beta}}
    \ev*{
    c_{\bm{k}+\bm{q}\alpha'}^{\dag}c_{\bm{k}+\bm{q}\beta}
    }
    \bra{\psi_{\bm{k}+\bm{q}\alpha'}}
    e^{i\bm{q}\cdot\bm{r}_2}
    \ket{\psi_{\bm{k}\beta'}}
    c_{\bm{k}\alpha}^{\dag}c_{\bm{k}\beta'}.
\end{align}
Here, by simplifying the Hartree term in matrix form, we get
\begin{align}
    \hat{H}_{\rm{H}}
    &=\sum_{\bm{k}}
    \sum_{\alpha,\beta}
    [h_{\rm{H}}(\bm{k})]_{\alpha\beta}
    c_{\bm{k}\alpha}^{\dag}c_{\bm{k}\beta},\notag
    \\
    h_{\rm{H}}(\bm{k})
    &=\frac{1}{A}
    \sum_{\bm{G}^M}
    V_{\bm{G}^M}
    \Lambda_{\bm{G}^M}(\bm{k})
    \sum_{\bm{k}'}
    \bigg(
    \operatorname{Tr}[
    P(\bm{k}')
    \Lambda_{\bm{G}^M}(\bm{k}')^*
    ]\bigg),
    \label{eq_Hart_app}
\end{align}
where single-particle density matrices and form factors 
are defined as
\begin{align}
[P(\bm{k})]_{\alpha\beta}
&=\ev*{c_{\bm{k}\alpha}^{\dag}c_{\bm{k}\beta}},
\label{eq:pmat}
\\
 [\Lambda_{\bm{q}}(\bm{k})]_{\alpha\beta}
&=\bra{\psi_{\bm{k}\alpha}}
e^{-i\bm{q}\cdot\bm{r}}
\ket{\psi_{\bm{k}+\bm{q}\beta}}.
\end{align}
Similarly, the Fock term is given by
\begin{align}
    \hat{H}_{\rm{F}}
    &=
    \sum_{\bm{k}}
    \sum_{\alpha,\beta}
    [h_{\rm{F}}(\bm{k})]_{\alpha\beta}
    c_{\bm{k}\alpha}^{\dag}c_{\bm{k}\beta},\notag
    \\
    h_{\rm{F}}(\bm{k})
    &=-\frac{1}{A}
    \sum_{\bm{q}}
    V_{\bm{q}}
    \Lambda_{\bm{q}}(\bm{k})
    P^T(\bm{k}+\bm{q})
    \Lambda_{\bm{q}}(\bm{k})^{\dag}.
    \label{eq_Fock_app}
\end{align}

In the HF calculation, the Schrödinger equation 
\begin{align}
    \sum_{\beta}
    (E^{(0)}_{\bm{k}\beta}\delta_{\alpha\beta} 
    + [h_{\rm{H}}(\bm{k})]_{\alpha\beta}
    &+ [h_{\rm{F}}(\bm{k})]_{\alpha\beta})
    u_{\bm{k}n}^{\beta}
    = E_{\bm{k}n}^{(\rm{HF})}u^{\alpha}_{\bm{k}n}
\end{align}
was solved self-consistently for each wavevector $\bm{k}$ 
using the single-particle eigenstates as basis.
Where $u^{\alpha}_{\bm{k}n}$ is a complex number, and the HF eigenvectors are given by
\begin{align}
    \ket{\psi^{\rm{HF}}_{\bm{k}n}}
     =\sum_{\alpha}u^{\alpha}_{\bm{k}n}
     \ket{\psi_{\bm{k}\alpha}}.
\end{align}

In addition, total energy $E_{\rm{tot}}$ can be evaluated by
\begin{align}
    E_{\rm{tot}}=
    \frac{1}{A}
    \sum_{\bm{k}}
    \operatorname{Tr}\bigg[
    \bigg(h_{0} (\bm{k})+ 
    \frac{h_{\rm{H}}(\bm{k})+h_{\rm{F}}(\bm{k})}{2}\bigg)^{\rm{T}} P(\bm{k})
    \bigg],
\end{align}
where $h_{0} (\bm{k})$ is a diagonal single-particle energy matrix
\begin{align}
    [h_{0} (\bm{k})]_{\alpha\beta}=
    E^{(0)}_{\bm{k}\alpha}\delta_{\alpha\beta}\ .
\end{align}

The Hartree term $h_{\rm{H}}(\bm{k})$, the Fock term $h_{\rm{F}}(\bm{k})$, the density matrix $P(\bm{k})$, and the form factors $\Lambda_{\bm{q}}(\bm{k})$ are all defined within the first moiré Brillouin zone and periodic with respect to $\bm{k}$. 
In contrast, the wave vector $\bm{q}$ of $\Lambda_{\bm{q}}(\bm{k})$ is not restricted to the first Brillouin zone; it must be specified for each $\bm{q}$ vector across distant Brillouin zones.
In our numerical implementation, we divide each moir\'{e} Brillouin zone into a uniform $24 \times 24$ $\bm{k}$-point mesh. 
In Eq.~\eqref{eq_Hart_app},~\eqref{eq_Fock_app}, the summation over $\bm{G}^M$ is restricted to 19 moiré Brillouin zones defined by $\bm{G}^M = n_1 \bm{G}^M_1 + n_2 \bm{G}^M_2$ with $|n_1|, |n_2| \leq 2$, while the summation over $\bm{q}$ is carried out for each $\bm{q}$ vector at every $\bm{k}$-point in the $24 \times 24$ mesh covering all these zones.

For the Hartree–Fock calculations, we initially project the interaction onto the lowest seven conduction bands for each spin and valley, neglecting the valence bands. 
This approximation is expected to be appropriate when the single-particle gap induced by the displacement field $u_D$ is sufficiently large~\cite{dong2024anomalous}. 
The validity of this projection scheme, including the dependence on the number of conduction bands retained and the effect of explicitly including valence bands, is examined in detail in the following section.
Because the Hartree-Fock equations may admit multiple self-consistent solutions corresponding to different symmetry sectors, we explore possible ground states at filling $\nu = 2$ by performing self-consistent calculations starting from different initial conditions for the single-particle density matrix $P(\bm{k})$.

In particular, the non-Abelian state is obtained by assuming valley polarization and allowing for spin-off-diagonal components in $P(\bm{k})$.
The quantum spin Hall (QSH) state is realized by constraining the spin-off-diagonal components of $P(\bm{k})$ to zero and initializing the occupation such that one electron occupies the valence band in each of the $K$ and $K'$ valleys.
A metallic solution is obtained by assuming valley polarization and a continuous filling of the conduction bands without opening a gap.
For each case, the initial $P(\bm{k})$ is taken to be a small random complex matrix, without imposing any additional bias toward a particular self-consistent solution. 
The self-consistent iterations are continued until $P(\bm{k})$ converges.
After convergence, the Hartree-Fock ground state at each set of parameters is identified as the self-consistent solution with the lowest total energy.

\section{Dependence of the ground state on band projection}

\begin{figure}
 \centering
\includegraphics[width=0.7\linewidth]{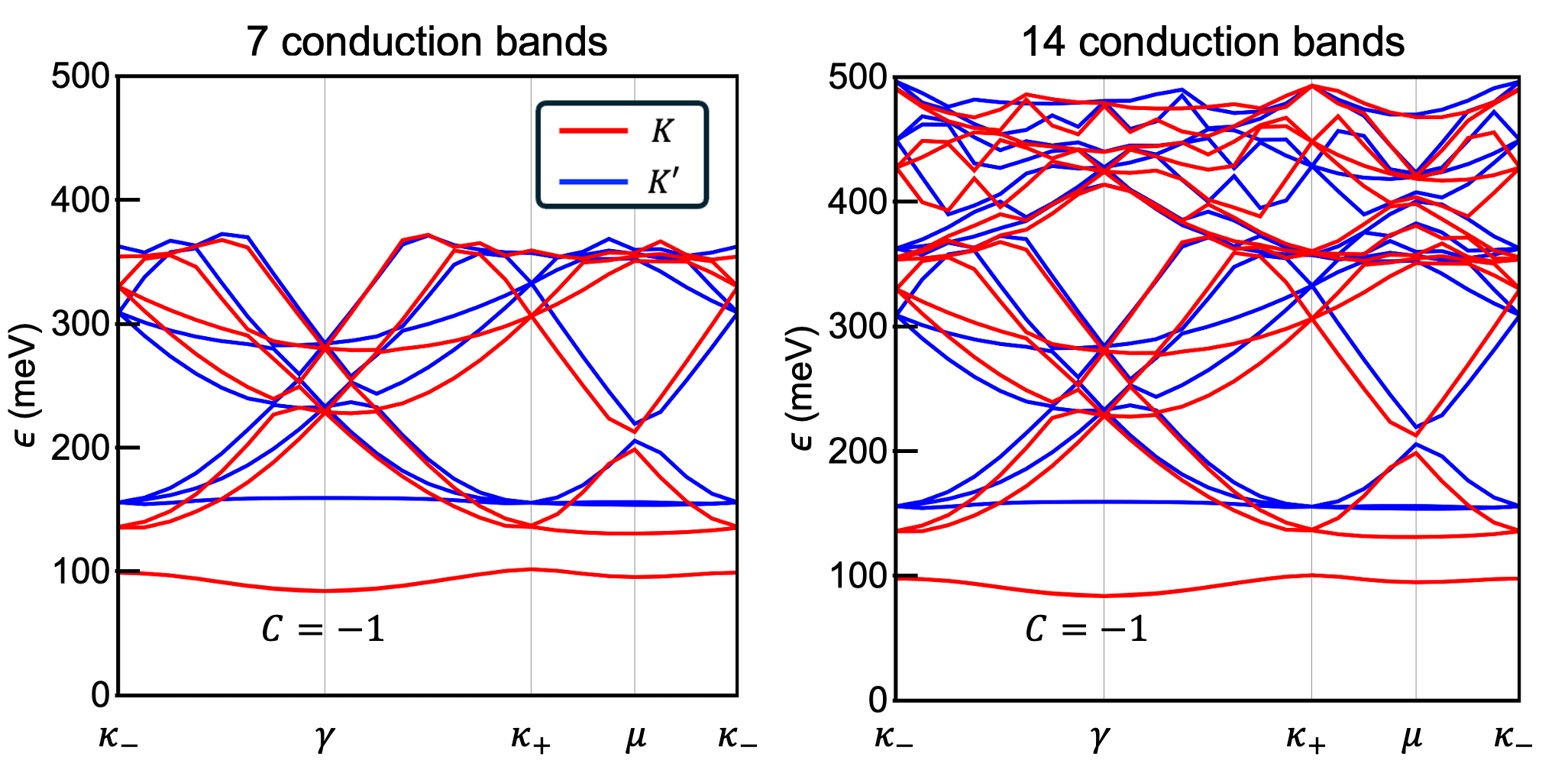}
    \caption{Band structures of a 5-layer system without hBN at $u_D = 40$~meV and $\theta = 0.77^\circ$, 
    comparing HF calculations with 7 (left panel) and 14 (right panel) conduction bands per spin and valley.
   }
    \label{fig:Condeuctionbands}
\end{figure}

\begin{figure}
\centering
\includegraphics[width=0.4\linewidth]{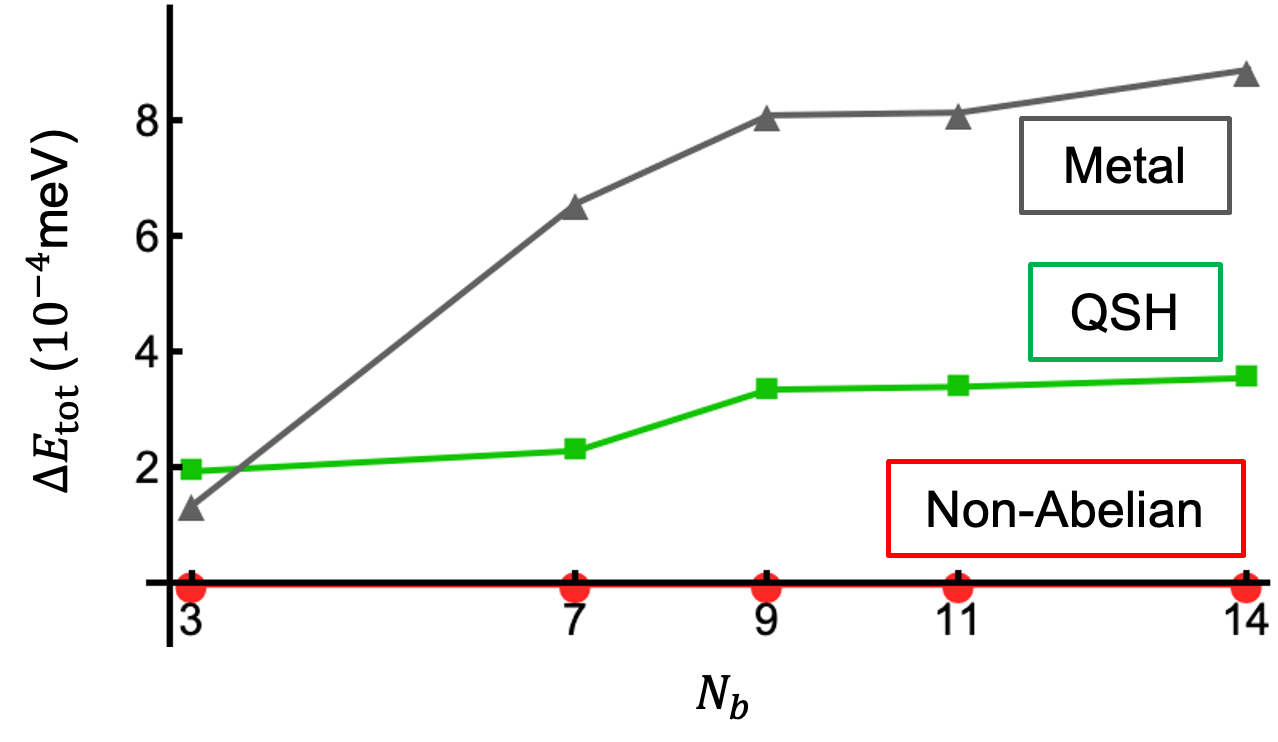}
    \caption{Total energy differences relative to the non-Abelian HF ground state as a function of the number of conduction bands $N_b$ per spin and valley (5-layer without hBN, $u_D = 40$ meV, $\theta = 0.77^\circ$).
    The plotted $\Delta E_{\rm{tot}}$ values show systematic convergence with increasing $N_b$, while the quantum spin Hall (QSH) and metallic states remain energetically higher for all $N_b$.
    From these results, we conclude that $N_b = 7$ is sufficient to reliably distinguish the ground state.}
    \label{fig:Nconduction}
\end{figure}

In the Hartree–Fock calculation in the main text, we project the continuum Hamiltonian onto the lowest seven conduction bands for each spin and valley in order to determine the ground state at $\nu = 2$.  
In the following, we examine the dependence of the ground state on the choice of projected bands and demonstrate the robustness of the emergent non-Abelian state.

We first increase the number of conduction bands $N_b$ included in the projection.  
Figure~\ref{fig:Condeuctionbands} shows the ground-state band structures of a 5-layer system without hBN at $u_D = 40$~meV and $\theta = 0.77^\circ$, obtained using seven (left panel) and fourteen (right panel) conduction bands per spin and valley.  
In both cases, the ground state is identified as a non-Abelian state, and the resulting band structures exhibit negligible differences, indicating convergence with respect to the number of conduction bands.
In Fig.~\ref{fig:Nconduction}, we present the total energy differences relative to the non-Abelian Hartree–Fock ground state as a function of the number of conduction bands $N_b$ per spin and valley (5-layer system without hBN, $u_D = 40$~meV, $\theta = 0.77^\circ$).  
The energetic hierarchy among competing candidate states remains unchanged as $N_b$ is varied.  
Moreover, the energy separation between the non-Abelian state and other competing states increases with increasing $N_b$, indicating that the non-Abelian phase becomes increasingly energetically favorable as more bands are included.

\begin{figure}
    \centering
    \includegraphics[width=0.7\linewidth]{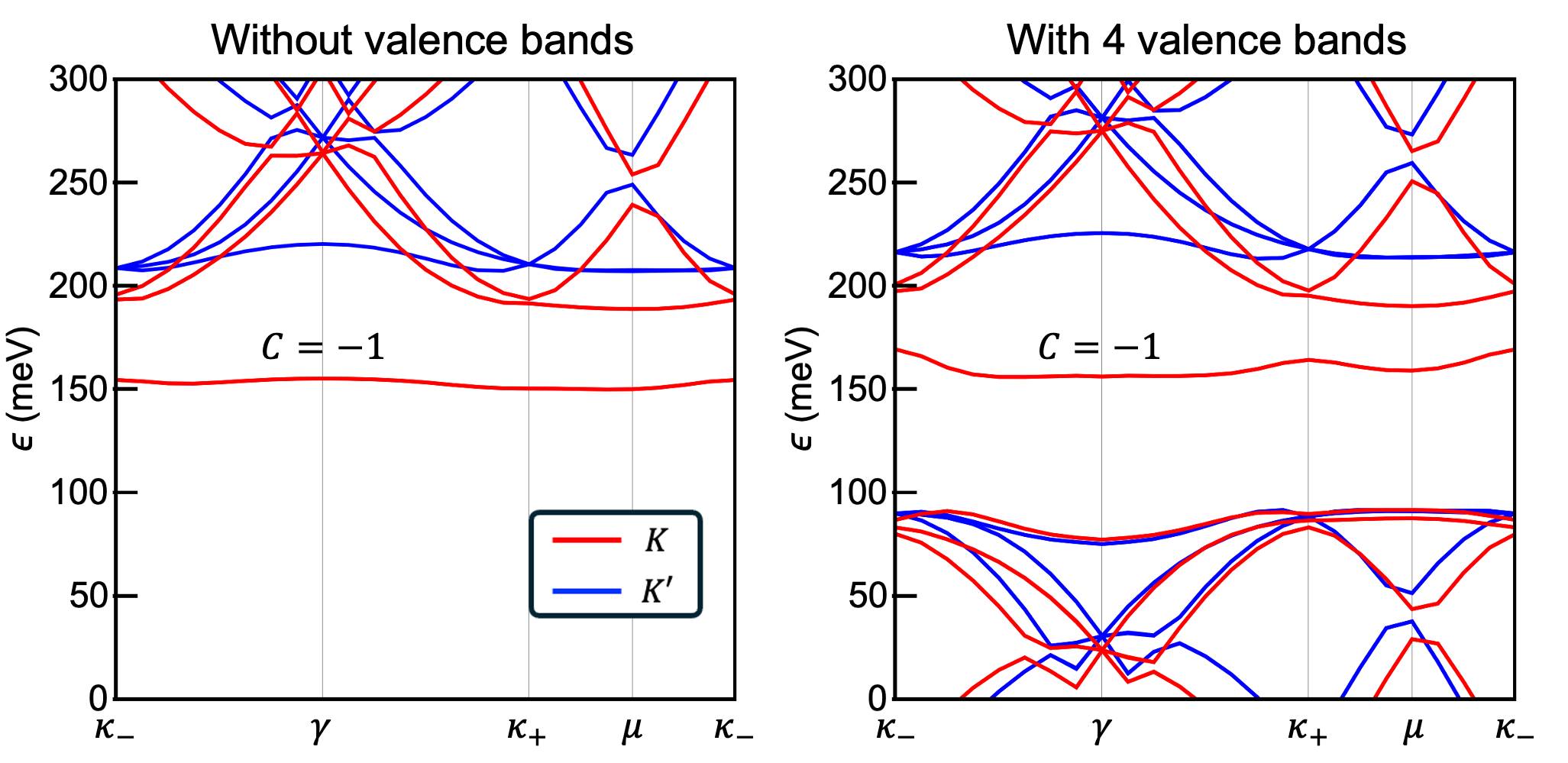}
    \caption{
    Non-Abelian Hartree–Fock solutions for a 3-layer system without hBN at $u_D = 110$~meV and $\theta = 0.4^\circ$.  
    The left panel shows results obtained by including only seven conduction bands, while the right panel includes four valence bands in addition to seven conduction bands.
    }
    \label{fig:WithValence}
\end{figure}

We also assess the sensitivity of the HF ground state to the inclusion of valence bands in the calculation.  
To incorporate the valence bands, the density matrix $P(\bm{k})$ [Eq.~\eqref{eq:pmat}] is modified by adding a reference contribution $P_{\rm Ref}(\bm{k})$~\cite{kwan2025moire}.  
The matrix elements of $P_{\rm Ref}(\bm{k})$ are defined as
\begin{align}
    [P_{\rm Ref}(\bm{k})]_{m s \xi,\, n s' \xi'} =
    \begin{cases}
        \delta_{mn}\,\delta_{ss'}\,\delta_{\xi\xi'}, & \text{if $m$ is a valence band},\\
        0, & \text{otherwise},
    \end{cases}
\end{align}
where $m,n$ denote band indices, and $(s,s')$ and $(\xi,\xi')$ label the spin and valley degrees of freedom, respectively.

Figure~\ref{fig:WithValence} compares the non-Abelian HF solutions for a 3-layer system without hBN at $u_D = 110$~meV and $\theta = 0.4^\circ$.  
The left panel shows results obtained by including only seven conduction bands, while the right panel includes four valence bands in addition to seven conduction bands.  
In both cases, the non-Abelian solutions exhibit qualitatively similar band structures.  
In the latter case, the non-Abelian state is energetically favored over both the QSH and metallic states, with total energy differences of approximately  
$\Delta E_{\rm tot} \simeq 8.0 \times 10^{-4}$~meV relative to the QSH state and  
$\Delta E_{\rm tot} \simeq 2.8 \times 10^{-3}$~meV relative to the metallic state.

\section{Details of the ground states at $\nu=2$}

\begin{figure}
    \centering
    \includegraphics[width=\linewidth]{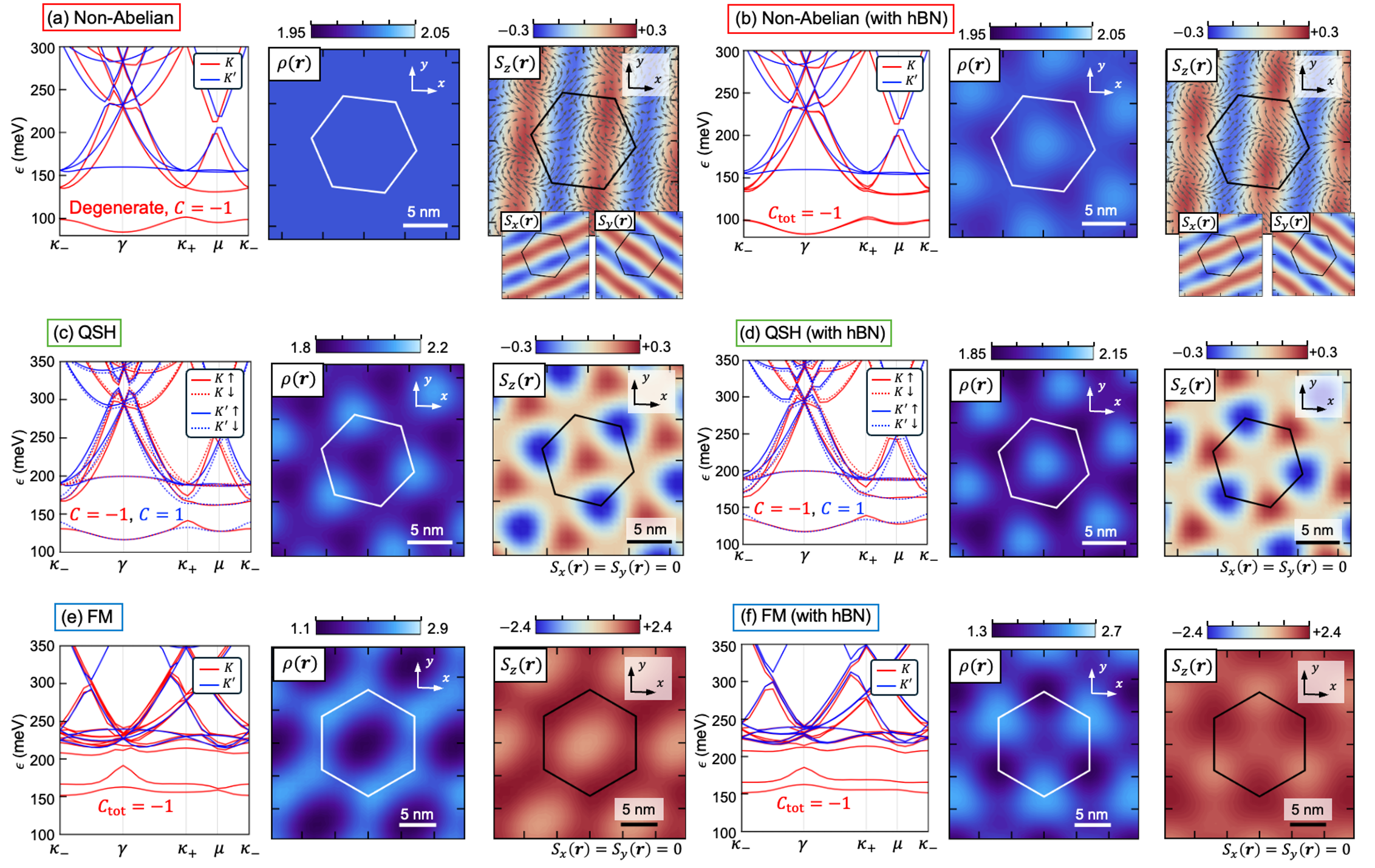}
    \caption{Band structures, charge distributions, and spin textures of the ground states corresponding to the phase diagram (Fig.  
    1 in the main text).
    (a) Non-Abelian state (5-layer, $u_D=40$ meV, $\theta=0.77^\circ$); (b) same parameter point as (a), but with hBN.
    (c) QSH state (5-layer, $u_D=50$ meV, $\theta=1^\circ$); (d) same parameter point as (c), but with hBN.
    (e) Ferromagnetic state (5-layer, $u_D=60$ meV, $\theta=0^\circ$); (f) same parameter point as (e), but with hBN.}
    \label{fig:ground}
\end{figure}

Fig.~\ref{fig:ground} shows the ground-state bands, charge distribution, and spin textures corresponding to the phases identified in the phase diagram (Fig. 
1 in the main text).
As shown in Fig.~\ref{fig:ground}(a) the non-Abelian state is valley-polarized and possesses a spin-degenerate Chern band with $C=-1$ and a spin texture characterized by a magnetic winding number of $+2$.
In the $K$ valley, these values are $C=-1$ and winding number $+2$, whereas in the $K'$ valley, both the Chern number and the winding number take the opposite sign.
In this state, the charge distribution is nearly uniform.
The spin textures $S_x$, $S_y$, and $S_z$ each exhibit cosine-like modulations along different translation vectors. 
In the presence of hBN [Fig.~\ref{fig:ground}(b)],the non-Abelian Chern band is slightly lifted, and the charge distribution shows slight modulation due to the hBN moir\'{e} potential.
Meanwhile, the spin textures $S_x$, $S_y$, and $S_z$ retain their cosine-like modulations with a magnetic winding number of $+2$.
Fig.~\ref{fig:ground}(c) and (d) show the quantum spin Hall (QSH) state.
In this state, each of the two valleys hosts a spin-polarized $|C|=1$ anomalous Hall crystal. 
The Chern numbers of the Chern bands in the two valleys have opposite signs, so the total Chern number is zero.
The spins are aligned antiparallel between the valleys.
Fig.~\ref{fig:ground}(d) and (e) show the ferromagnetic state.
In this state, both valleys and spins are polarized, the Chern bands are non-degenerate, and the total Chern number is $|C|=1$.
Without hBN, the charge distribution forms a slightly distorted hexagon and lacks threefold rotational symmetry. 
In contrast, with hBN, the charge distribution exhibits threefold rotational symmetry.

\begin{figure}
    \centering
    \includegraphics[width=0.4\linewidth]{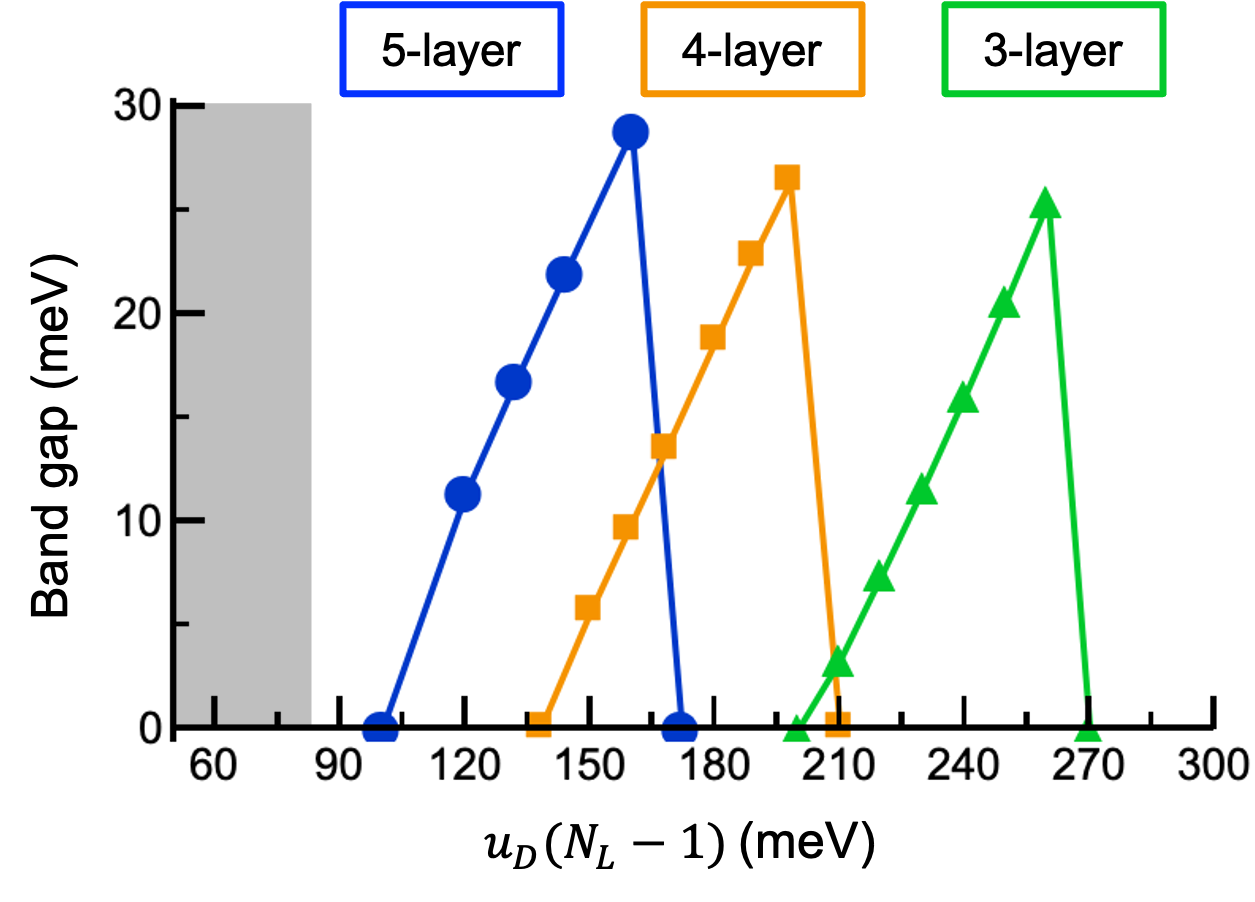}
    \caption{Band gap for rhombohedral 3-, 4- and 5-layer graphene in the absence of hBN. 
    The electronic period $L^M$ is fixed at 11.15 nm ($\theta=0.77^\circ$).
    All gapped ground states are non-Abelian states. The gray area indicates the regime where the electric field is too weak for the valence bands to be neglected. In this gray region, our Hartree-Fock calculations are not accurate.
    }
    \label{fig:gap_nohBN}
\end{figure}

\begin{figure}
    \centering
    \includegraphics[width=0.7\linewidth]{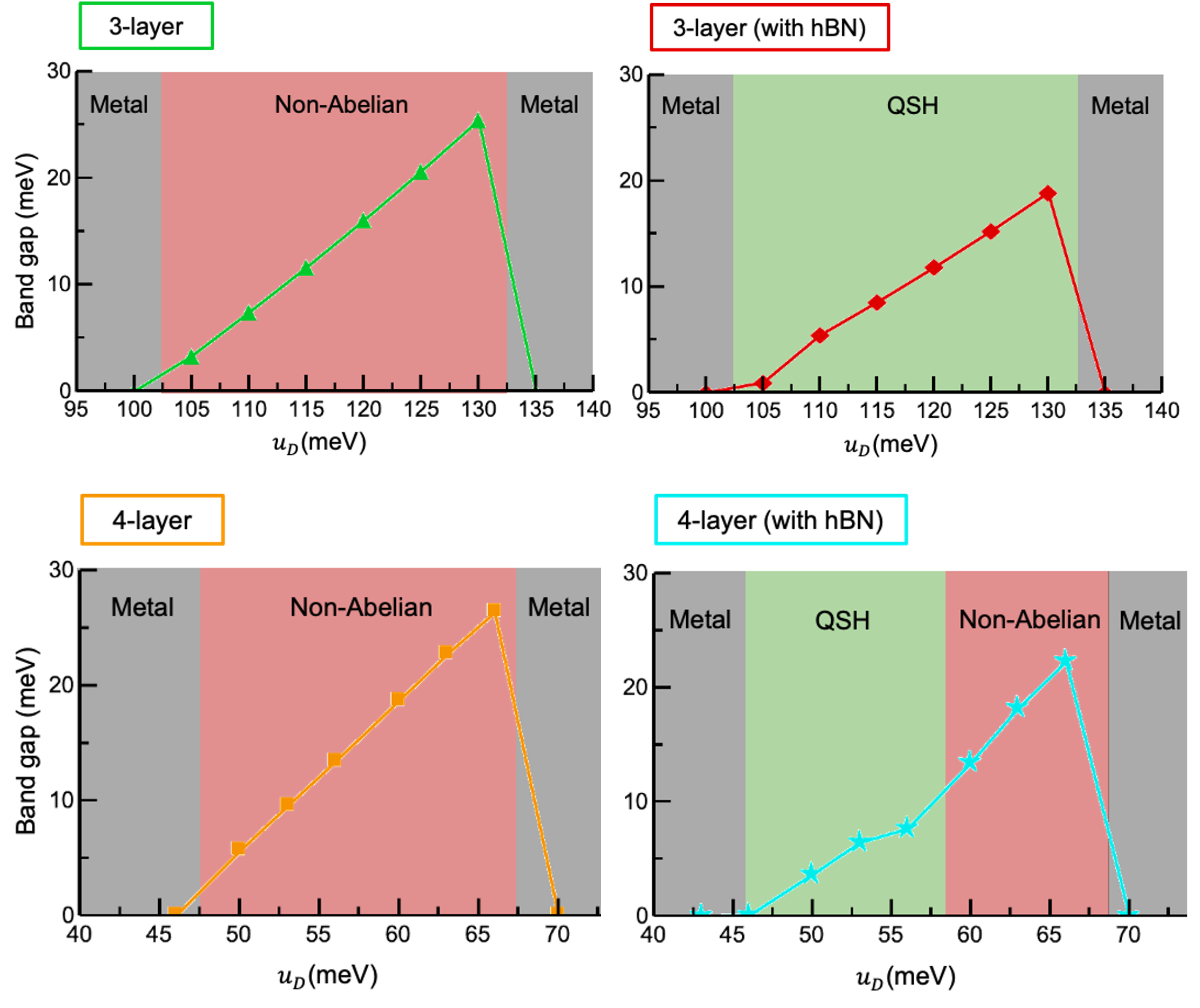}
    \caption{Band gap for 3- and 4-layer systems at $L^M=11.15$ nm ($\theta=0.77^\circ$). The left panels show the cases without hBN, and the right panels show the cases with hBN. 
    If the influence of $u_D$ is smaller than that of the hBN potential, the non-Abelian state gives way to the QSH state.}
    \label{fig:gap_hBN}
\end{figure}

We now discuss the band gap. Fig.~\ref{fig:gap_nohBN} shows the band gaps of the ground states in 3-, 4-, and 5-layer graphene at each perpendicular electric field $u_D$ in the absence of hBN. The electronic period $L^M$ is fixed at 11.15 nm ($\theta=0.77^\circ$).
From Fig.~\ref{fig:gap_nohBN}, we observe that the band gap increases almost linearly with increasing $u_D$.
As the number of layers $N_L$ increases, the $u_D$ required for the emergence of the non-Abelian state decreases. 
Similarly, the $u_D$ at which the metallic state appears also decreases.
However, when the $u_D$ becomes too small, the gap between the conduction and valence bands becomes very small or closes entirely, making it likely that a different phase emerges.
We have not explicitly confirmed this, because in such cases, our HF calculations that focus only on the conduction bands may no longer be reliable.
Therefore, the non-Abelian state does not necessarily become more stable with increasing number of layers.
Fig.~\ref{fig:gap_hBN} shows the band gaps of the ground states in 3- and 4-layer systems with and without hBN.
In this work, we focus on electronic states localized away from the twisted hBN interface.
When the number of layers is small, the hBN effect becomes more pronounced because the electronic states are located closer to the hBN.
However, increasing the interlayer potential difference $u_D$, induced by a perpendicular electric field, enhances the layer polarization of the conduction band states, pushing them further away from the hBN interface and thereby reducing its impact.
As a result, when hBN is present and either $u_D$ is small or the number of layers is low, the hBN effect becomes strong, and the non-Abelian state transitions to a QSH state.
These trends are clearly observed in Fig.~\ref{fig:gap_hBN}.

\end{document}